\documentclass{elsart}
\usepackage[dvips]{graphicx}
\newcommand{\krzb}{\ensuremath{\overline{K}^{*0}}}
\newcommand{\krzmndk}{\ensuremath{D^+ \rightarrow \krzb \mu^+ \nu}}
\newcommand{\krzlndk}{\ensuremath{D^+ \rightarrow \krzb \ell^+ \nu_\ell}}
\newcommand{\philndk}{\ensuremath{D_s^+ \rightarrow \phi\; \ell^+ \nu_\ell}}
\newcommand{\phimndk}{\ensuremath{D_s^+ \rightarrow \phi\; \mu^+ \nu }}
\newcommand{\kkmndk}{\ensuremath{D_s^+ \rightarrow K^+ K^- \mu^+ \nu }}

\newcommand{\kpimndk}{\ensuremath{D^+ \rightarrow K^- \pi^+ \mu^+ \nu }}

\newcommand{\gevcsq}{\ensuremath{\textrm{GeV}/c^2}}

\newcommand{\thv}{\ensuremath{\theta_\textrm{v}}}
\newcommand{\thl}{\ensuremath{\theta_\ell}}
\newcommand{\costhv}{\ensuremath{\cos\thv}}

\newcommand{\costhl}{\ensuremath{\cos\thl}}

\newcommand{\qsq}{\ensuremath{q^2}}
\newcommand{\yvar}{\ensuremath{\qsq{}/\qsq_{\rm max}}}
\newcommand{\bw}{\ensuremath{\textrm{B}_{\phi}}}
\newcommand{\mkpi}{\ensuremath{m_{K\pi}}}
\newcommand{\mkk}{\ensuremath{m_{K^+ K^-}}}

\newcommand{\gevc}{\ensuremath{\textrm{GeV}^2/c^2}}

\newcommand{\rvvalue}{\ensuremath{1.549 \pm 0.250 \pm 0.145}}
\newcommand{\rtwovalue}{\ensuremath{0.713 \pm 0.202 \pm 0.266}}
\newcommand{\rvresult}{\ensuremath{r_v = \rvvalue{}}}
\newcommand{\rtworesult}{\ensuremath{r_2 = \rtwovalue{}}}
\newcommand{\rtwo}{\ensuremath{r_2}}
\newcommand{\rthree}{\ensuremath{r_3}}
\newcommand{\rvee}{\ensuremath{r_v}}

\newcommand{\mysection}[1]{\section{#1}}
\newcounter{saveeqn}%

\begin{document}
\begin{frontmatter}
\title{New Measurements of the \phimndk{} Form Factor Ratios}
%%%%%%% Do not change authors here.  Use the database!
\author[ucd]{J.~M.~Link}
\author[ucd]{P.~M.~Yager}
\author[cbpf]{J.~C.~Anjos}
\author[cbpf]{I.~Bediaga}
\author[cbpf]{C.~G\"obel}
\author[cbpf]{A.~A.~Machado}
\author[cbpf]{J.~Magnin}
\author[cbpf]{A.~Massafferri}
\author[cbpf]{J.~M.~de~Miranda}
\author[cbpf]{I.~M.~Pepe}
% ...

\address[ucd]{University of California, Davis, CA 95616} 
\address[cbpf]{Centro Brasileiro de Pesquisas F\'\i sicas, Rio de Janeiro, RJ, Brasil} 
\address[cinv]{CINVESTAV, 07000 M\'exico City, DF, Mexico} 
\address[cu]{University of Colorado, Boulder, CO 80309} 
\address[fnal]{Fermi National Accelerator Laboratory, Batavia, IL 60510} 
\address[fras]{Laboratori Nazionali di Frascati dell'INFN, Frascati, Italy I-00044}
\address[guan]{University of Guanajuato, 37150 Leon, Guanajuato, Mexico} 
\address[ui]{University of Illinois, Urbana-Champaign, IL 61801} 
\address[iu]{Indiana University, Bloomington, IN 47405} 
\address[korea]{Korea University, Seoul, Korea 136-701}
\address[korea2]{Kyungpook National University, Taegu, Korea 702-701}
\address[milan]{INFN and University of Milano, Milano, Italy} 
\address[nc]{University of North Carolina, Asheville, NC 28804} 
\address[pavia]{Dipartimento di Fisica Nucleare e Teorica and INFN, Pavia, Italy} 
\address[pr]{University of Puerto Rico, Mayaguez, PR 00681} 
\address[sc]{University of South Carolina, Columbia, SC 29208} 
\address[ut]{University of Tennessee, Knoxville, TN 37996} 
\address[vu]{Vanderbilt University, Nashville, TN 37235} 
\address[wisc]{University of Wisconsin, Madison, WI 53706}

\endnote{\small See http://www-focus.fnal.gov/authors.html for
additional author information}
\nobreak
\begin{abstract}
Using a large sample of \phimndk{} decays
collected by the FOCUS photoproduction experiment at Fermilab, we
present new measurements of two semileptonic form factor ratios: \rvee{}
and \rtwo{}.  We find \rvresult{} and \rtworesult{}. These values
are consistent with \rvee{} and \rtwo{} form factors measured for the 
process \krzlndk{}.
\end{abstract}
\end{frontmatter}
\newpage
% No page number printed for this page
%\tableofcontents    % contents are based on the sections, subsections, etc.
%\listoffigures     % based on the figures
\newpage

\mysection{Introduction}

This paper provides new measurements of the parameters that
describe \phimndk{} decay.  The \phimndk{} decay amplitude
is described~\cite{KS} by four form factors with an assumed (pole form) \qsq{}
dependence.  Following earlier experimental work~
\cite{focus_ff,anomaly,beatrice,e791e,e791mu,e791phi,cleophi,e687,e687phi,e653,e653phi,e691},
the \phimndk{} amplitude is then described by ratios of form factors taken
at \qsq{} = 0. The traditional set is: \rtwo{}, \rthree{}, and \rvee{}
which we define explicitly after Equation \ref{amp1}.
According to flavor SU(3) symmetry, one expects that the 
the form factor ratios describing \phimndk{} should 
be close to those describing \krzmndk{} since the $D_s^+$ 
only differs from the $D^+$ through the replacement of a $\overline{d}$
quark by a $\overline{s}$ quark spectator.   
The existing lattice gauge calculations \cite{theory} predict that the form factor ratios describing \philndk{} should lie within 10\% of those describing \krzlndk{}.   
Although the measured \rvee{} form factors are quite consistent between \philndk{} and \krzlndk{},
there is presently a 3.3 $\sigma$ discrepancy between the \rtwo{} values measured
for these two processes with the previously measured \philndk{} value being a factor of about 1.8 times
larger than the \rtwo{} value measured for \krzlndk{}~\cite{e791phi}.  One quark model
calculation \cite{ISGW2} offers a possible explanation for the apparent inconsistency
in the \rtwo{} values measured for \krzlndk{} and \philndk{}.

\begin{figure}[tbph!]
 \begin{center}
  \includegraphics[width=3.0in]{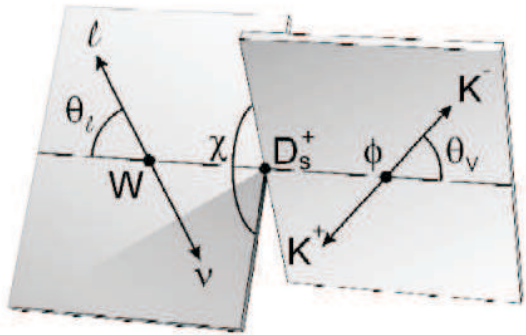}
  \caption{Definition of kinematic variables.
 \label{angles}}
 \end{center}
\end{figure}

Five kinematic variables that uniquely describe \kkmndk{} decay are
illustrated in Figure~\ref{angles}. These are the $K^- K^+$
invariant mass (\mkk{}) , the square of the $\mu\nu$ mass (\qsq{}),
and three decay angles: the angle between the $K^+$ and the $D_s^+$
direction in the $K^- K^+$ rest frame (\thv{}), the angle between
the $\nu$ and the $D_s^+$ direction in the $\mu\nu$ rest frame (\thl{}),
and the acoplanarity angle between the two decay planes
($\chi$). These angular conventions on \thl{} and \thv{} apply to
both the $D_s^+$ and $D_s^-$. The sense of the acoplanarity variable is
defined via a cross product expression of the form: $ (\vec P_\mu
\times \vec P_\nu) \times (\vec P_{K^-} \times \vec P_{K^+}) \cdot \vec P_{{K^-}
{K^+}}$ where all momentum vectors are in the $D_s^+$ rest frame.  Since
this expression involves five momentum vectors, as one goes from $D_s^+
\rightarrow D_s^-$ one must change $\chi \rightarrow -\chi$ in
Equation~\ref{amp1} to get the same intensity for the $D_s^+$ and $D_s^-$ 
assuming \emph{CP} symmetry.

Using
the notation of \cite{KS}, we write the decay distribution for
\phimndk{} in terms of the four helicity basis form factors:
$H_+~,~H_0~,~H_-~,H_t$.  
\begin{eqnarray}
{d^5 \Gamma \over dm_{K \pi}~d\qsq~d\cos\thv~d\cos\thl~d\chi}
\propto K (\qsq{} - m_l^2 )\left\{\left| \begin{array}{l}
 (1 + \cos \thl )\sin \thv e^{i\chi } \bw H_ +   \\
  - \,(1 - \cos \thl )\sin \thv e^{-i\chi } \bw H_ -   \\
  - \,2\sin \thl (\cos \thv \bw + Ae^{i\delta } )H_0  \\
 \end{array} \right|^2 \right. \nonumber \\
 + \frac{m^2_\ell}{\qsq} \left. \left| \begin{array}{l}
 \sin\thl \sin\thv \bw \left(e^{i\chi}H_+ + e^{-i\chi}H_-\right) \\
+\,2\cos\thl\left(\cos\thv\bw + Ae^{i\delta}\right)H_0 \\
+\,2\left(cos\thv\bw + Ae^{i\delta}\right)H_t
 \end{array} \right|^2 \right\}\phantom{xxx}
\label{amp1}
\end{eqnarray}

where $K$ is the momentum of the $K^- K^+$ system in the rest frame
of the $D_s^+$. The first term gives the intensity for the $\mu^+$ to be
right-handed, while the (highly suppressed) second term gives the
intensity for it to be left-handed.  The helicity basis form factors
are given by:
\begin{eqnarray*}
H_\pm (\qsq) &=& (M_D+\mkpi)A_1(\qsq)\mp 2{M_D K\over M_D+m_{K\pi}}V(\qsq)\\
H_0 (\qsq) &=& {1\over 2\mkpi\sqrt{\qsq}}
\left[ (M^2_D -m^2_{K\pi}-\qsq)(M_D+\mkpi)A_1(\qsq)
-4{M^2_D K^2\over M_D+\mkpi}A_2(\qsq) \right] \\
H_t(\qsq) &=& {M_D K\over M_{K\pi}\sqrt{\qsq}}
\left[  (M_D+M_{K\pi})A_1(\qsq) - {(M^2_D -M^2_{K\pi}+\qsq) \over M_D+M_{K\pi}}A_2(\qsq)
+{2\qsq\over M_D+M_{K\pi}}A_3(\qsq) \right]
\end{eqnarray*}
The vector and axial form factors are generally parameterized by a pole
dominance form:
\[
A_i(\qsq)={A_i(0)\over 1-\qsq/M_A^2}~~~~~~~~
V(\qsq)={V(0)\over 1-\qsq/M_V^2}
\]
where we use nominal (spectroscopic) pole masses of $M_A =
2.5~\gevcsq$ and $M_V = 2.1~\gevcsq$.
The
\bw{} denotes the Breit-Wigner amplitude describing the $\phi$
resonance:\footnote {We are using a $p$-wave Breit-Wigner form with a
width proportional to the cube of the kaon momentum in the kaon-kaon
rest frame ($P^*$) over the value of this momentum when the kaon-kaon
mass equals the resonant mass ($P^*_0$).  The squared modulus of our
Breit-Wigner form will have an effective $P^{*3}$ dependence in the
numerator as well. Two powers $P^*$ come explicitly from the $P^*$ in
the numerator of the amplitude and one power arises from the 4 body
phase space.}
\[
\bw = \frac{\sqrt{m_0 \Gamma\,} \left(\frac{P^*}{P_0^*}\right)}
	   {m_{KK}^2 - m_0^2 + i m_0 \Gamma 
	\left(\frac{P^*}{P_0^*}\right)^3} 
\]

Equation \ref{amp1} includes a possible $s$-wave amplitude coupling to the virtual $W^+$ 
with the same \qsq{} dependence as that of the
$H_0$ (or $H_t$) form factor.  Evidence for such an $s$-wave amplitude
term for the decay \kpimndk{} was presented in Reference~\cite{anomaly}.  An explicit
search was made for $s$-wave amplitude interference with the process \phimndk{} and
no evidence for this interference was seen.  We were able to limit the $s$-wave
contribution to be less than 5\% of the maximum of the $\phi$ Breit-Wigner peak
in the $H_0$ piece of Equation \ref{amp1} at the 90\% confidence level.
The results presented here will therefore
assume $A = 0$ in Equation \ref{amp1}. Under these assumptions, the decay
intensity is then parameterized by the $\rvee{} \equiv V(0)/ A_1(0) ~,~
\rtwo{} \equiv A_2(0)/A_1(0)~,~\rthree{} \equiv A_3(0)/A_1(0)$ form factor 
ratios describing the \phimndk{} amplitude. 
Throughout this paper, unless explicitly stated otherwise,
the charge conjugate is also implied when a decay mode of a specific
charge is stated.

\mysection{Experimental and analysis details}

The data for this paper were collected in the Wideband photoproduction
experiment FOCUS during the Fermilab 1996--1997 fixed-target run. In
FOCUS, a forward multi-particle spectrometer is used to measure the
interactions of high energy photons on a segmented BeO target. The
FOCUS detector is a large aperture, fixed-target spectrometer with
excellent vertexing and particle identification. Most of the FOCUS
experiment and analysis techniques have been described
previously~\cite{anomaly,nim,ycp,CNIM,target}.
Our analysis cuts were chosen to give reasonably uniform acceptance
over the five kinematic decay variables, while maintaining a strong
rejection of backgrounds.  To suppress background from the
re-interaction of particles in the target region which can mimic a
decay vertex, we required that the charm secondary vertex was located
at least three  standard deviation outside of all solid
material including our target and target microstrip system.

To isolate the
\phimndk{} topology, we required that candidate muon, pion, and kaon
tracks appeared in a secondary vertex with a confidence level
exceeding 1\%.  The muon track, when extrapolated to the shielded muon
arrays, was required to match muon hits with a confidence level
exceeding 5\%. The kaon was required to have a \v Cerenkov light
pattern more consistent with that for a kaon than that for a pion by 1
unit of log likelihood \cite{CNIM}. To further reduce non-charm background we
required that our primary vertex consisted of at least two charged
tracks.  To further reduce muon misidentification, a muon candidate was allowed
to have at most one missing hit in the 6 planes comprising our inner
muon system and an energy exceeding 10 GeV.  In order to suppress
muons from pions and kaons decaying within our apparatus, we required
that each muon candidate had a confidence level exceeding 1\% to the
hypothesis that it had a consistent trajectory through our two
analysis magnets.

Non-charm and random combinatoric
backgrounds were reduced by requiring both a detachment between the
vertex containing the $K^-K^+\mu^+$ and the primary production
vertex of at least 5 standard deviations. 

\begin{figure}[tbph!]
 \begin{center}
\includegraphics[width=3.in]{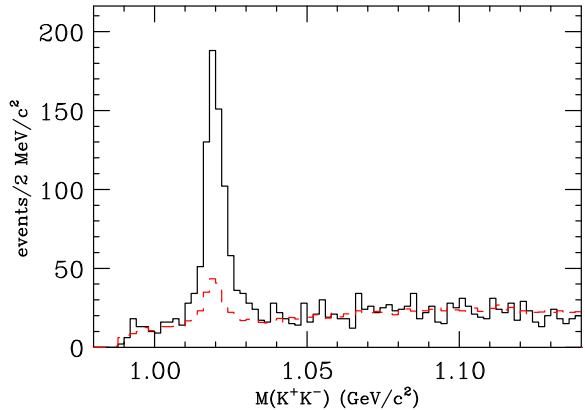}
\caption{ 
The data is the solid histogram and ccbar background Monte Carlo is
the dashed histogram. The ccbar background Monte Carlo is normalized to the same number
of events in the sideband region $1.040~\gevcsq{} < \mkk{} < 1.14~\gevcsq$. 
\label{signal}}
\end{center}
\end{figure}

Possible background from $D^+
\rightarrow K^- K^+ \pi^+$, where a pion is misidentified as a muon,
was reduced  by treating the muon as a pion and requiring the reconstructed $K K \pi$ mass
be less than  $1.8~\gevcsq{}$. The \mkk{} distribution for our \kkmndk{} candidates
is shown in Figure \ref{signal}.

The technique used to reconstruct the neutrino momentum through the
$D_s^+$ line-of-flight, and tests of our ability to simulate the
resolution on kinematic variables that rely on the neutrino momentum are
described in Reference~\cite{anomaly}.

\mysection{Fitting Technique} 

The \rvee{} and \rtwo{} form factors were fit to the probability density function described by
four fitted kinematic variables 
(\qsq{}, \costhv{}, \costhl{}, and $\chi$) for decays
in the mass range $1.010 < \mkk{} < 1.030$. 
We use a variant of the continuous fitting technique developed
by the E691 Collaboration~\cite{schmidt} for fitting decay intensities
where several of the kinematic variables have very poor resolution such
as the four variables that rely on reconstructed neutrino kinematics.

The fit which determines the \rvee{} and \rtwo{} form factor ratios minimizes the sum of $w = -2 \ln{I}$ where
$I$ is the normalized decay intensity at each datum. 
The intensity at each datum is estimated by the weight of Monte Carlo events that lie within
a small window of each of the four (reconstructed) kinematic variables of the given datum.\footnote{A Monte Carlo event had to lie within 0.08~to each datum in both \costhv{} and \costhl{}, within 0.18
radians in $\chi$, and within 0.8 \gevc{} in \qsq{}.} Some care is needed in choosing a reasonable size
of the windows since too small a window will result in fluctuations due to finite Monte Carlo statistics, and 
too large a window will create a bias in the result.  Our principal tool in deciding a reasonable window was 
to check for biases and fluctuations outside of reported 
statistical errors using a Monte Carlo simulation that included charm backgrounds with \rvee{} and \rtwo{} values
very close to the result reported here. Variations in the final results due to the window choice were included in the estimate of the systematic error.

Two Monte Carlos were employed: a weighted \phimndk{} signal Monte Carlo that was
generated flat in the four fitted kinematic variables (\qsq{}, \costhv{}, \costhl{}, and $\chi$) and an
unweighted background Monte Carlo which simulated all
known charm decays (apart from the \phimndk{} signal) as well as our misidentification levels.
The intensity about each datum is an appropriate average of the weighted
signal MC and the unweighted background Monte Carlo. The averaging depended on the background fraction determined
by matching the number of events in the $\phi$ sideband region $1.040 < \mkk{} < 1.14~\gevcsq{}$ 
in the background Monte Carlo to the sideband level observed in the data.

Our fit was to the \rvee{} and \rtwo{} form factor ratios with $\rthree{}$
and the possible $s$-wave amplitude set to zero. We decided not to fit for
the \rthree{} form factor ratio since our anticipated \rthree{} error
given our sample size would be too large to be meaningful.

\begin{figure}[htp]
\includegraphics[width=5in]{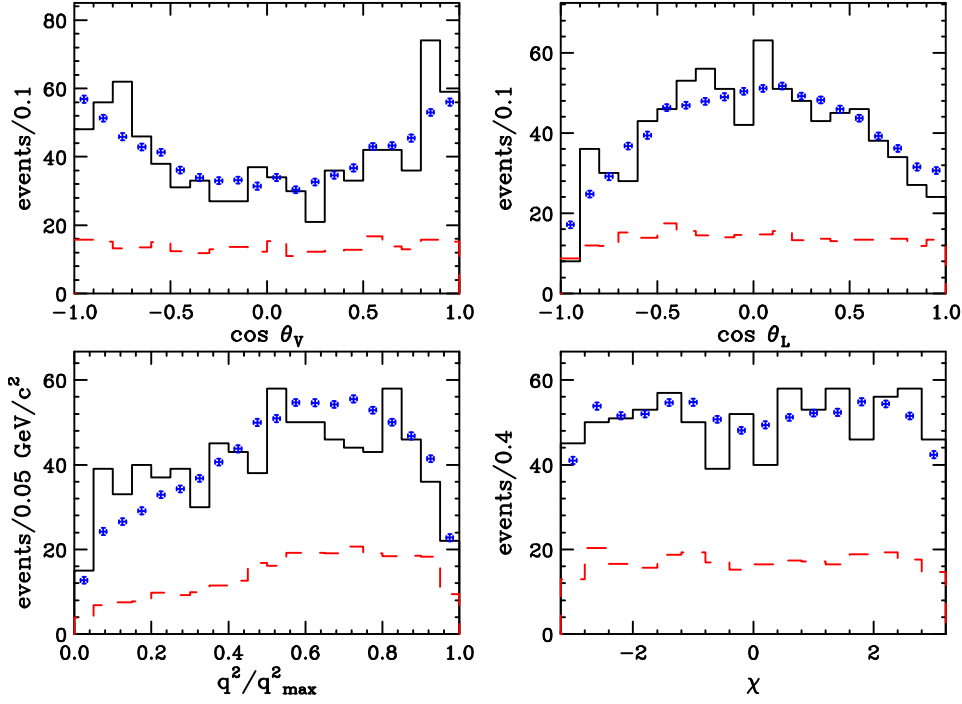}
\caption{ 
The data are given by the upper histogram.
The model (crosses with flats) includes the signal computed with the fitted form factor
ratios plus the sideband normalized ccbar background. The lower histogram (dashed) shows
the projections of the ccbar background. 
(a) \costhv{} 
(b) \costhl{}
(c) \yvar{}  
(d) $\chi$
\label{proj}}
\end{figure}

Figure \ref{proj} compares the data and model for
projections of \costhv{}, \costhl{}, $\chi$ and \yvar{}. Most
of these distributions follow the predicted values reasonably well.
A slight discrepancy is evident in the low \yvar{} projection
(below $\yvar{} < 0.2~$). A stronger effect was observed
in Reference \cite{focus_ff} possibly owing to our much larger yield in the \kpimndk{} final state.

\begin{figure}[htp]
\begin{center}
\includegraphics[width=2.5in]{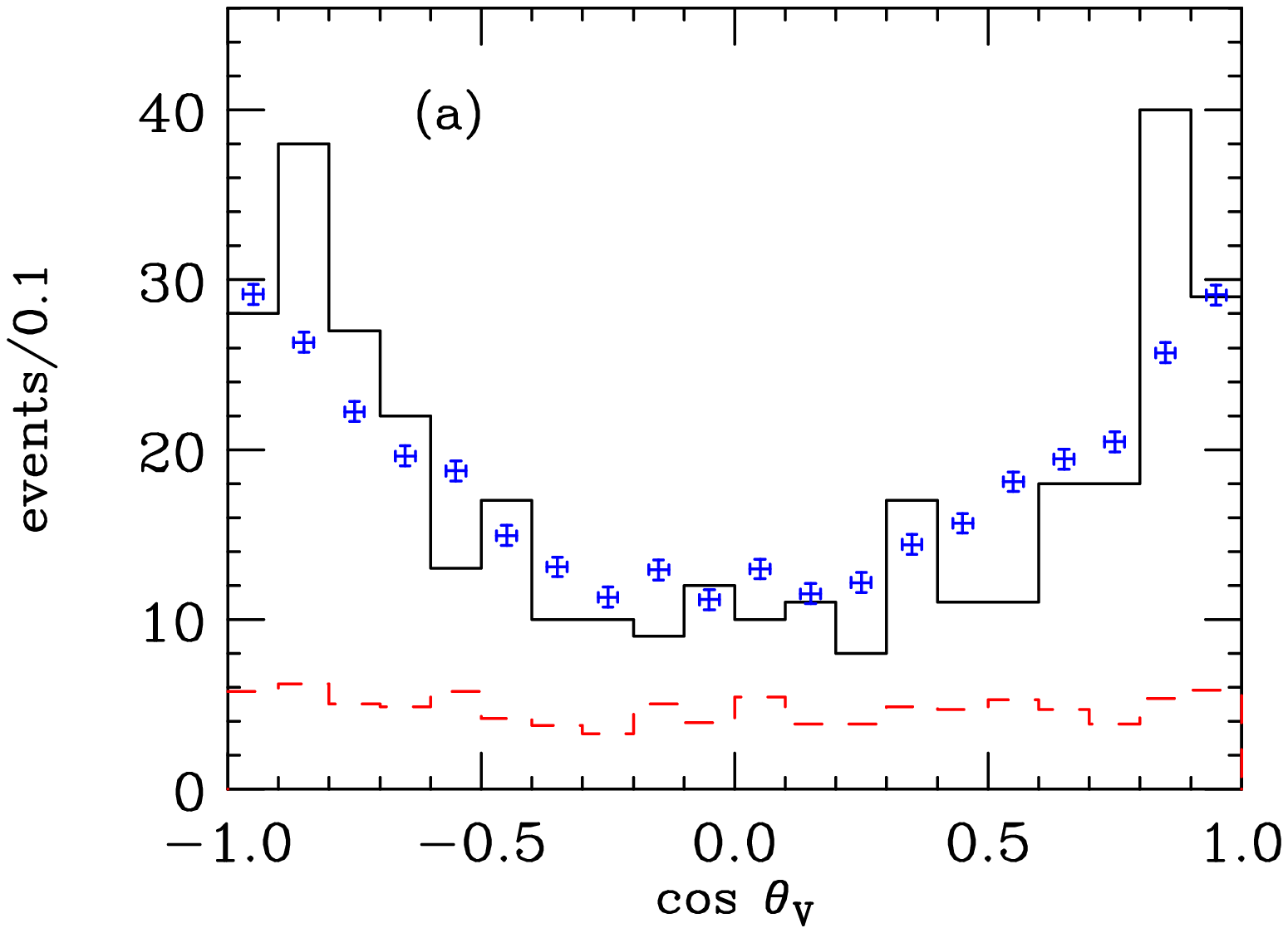}
\includegraphics[width=2.5in]{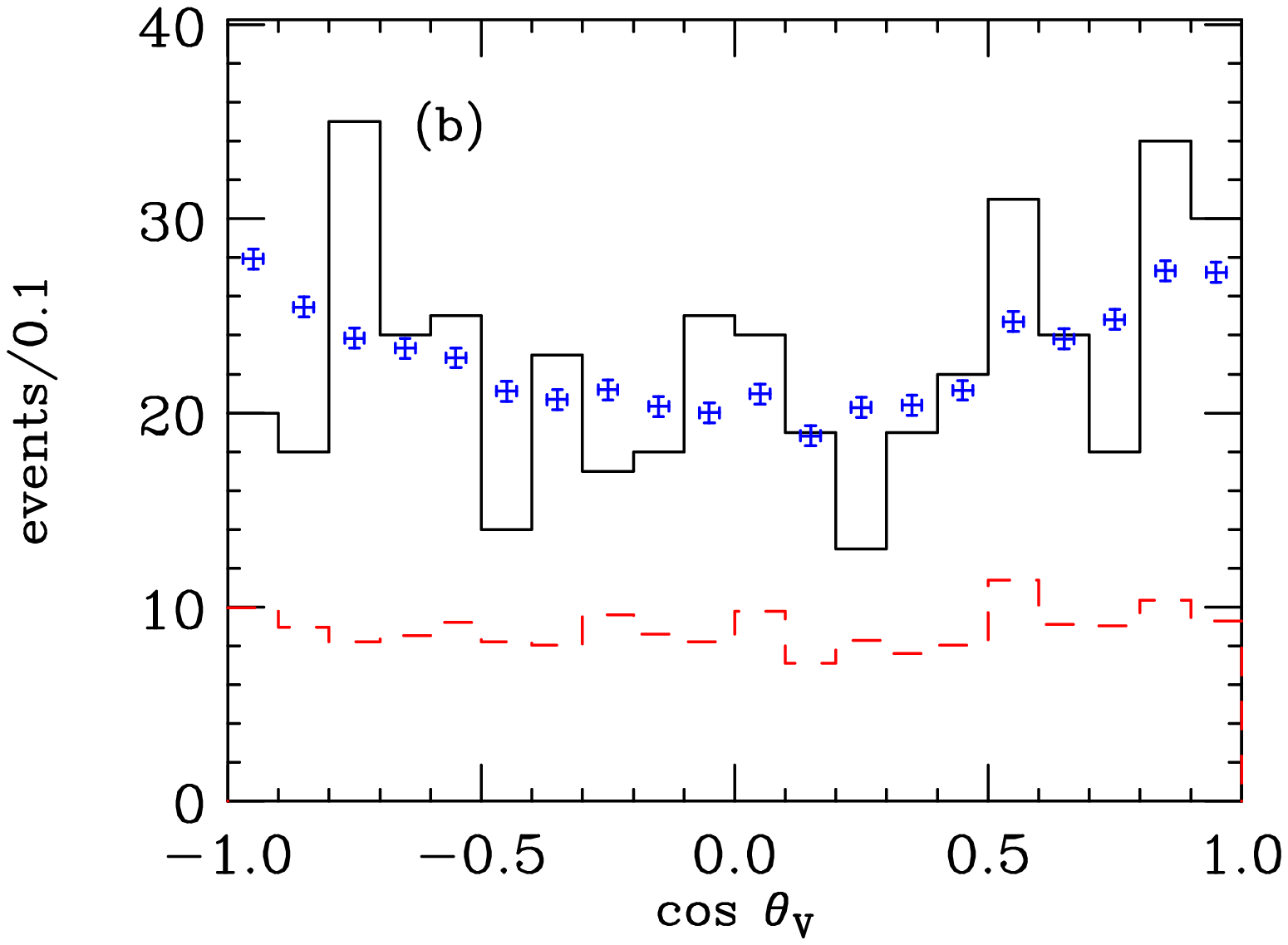}
\includegraphics[width=2.5in]{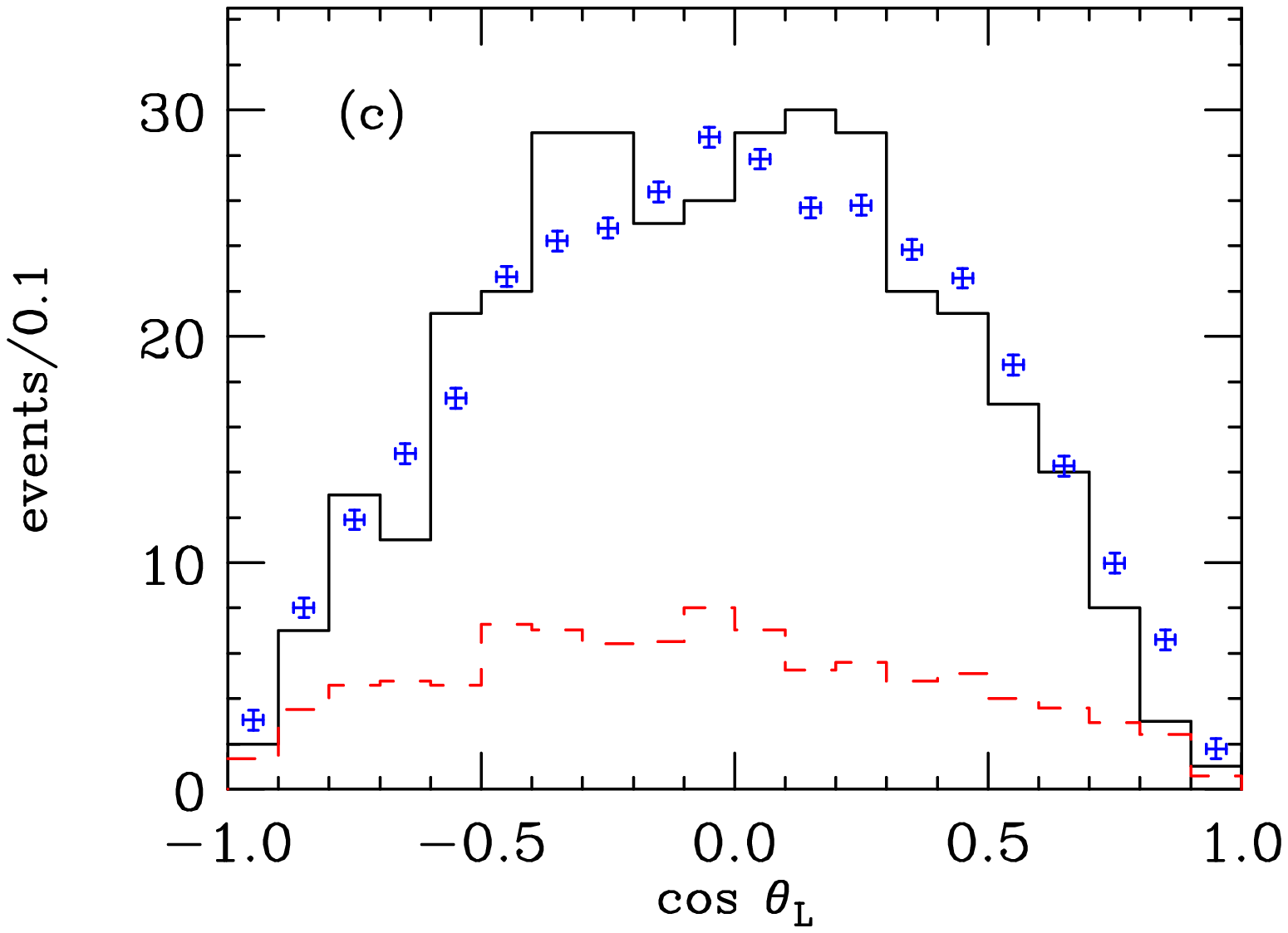}
\includegraphics[width=2.5in]{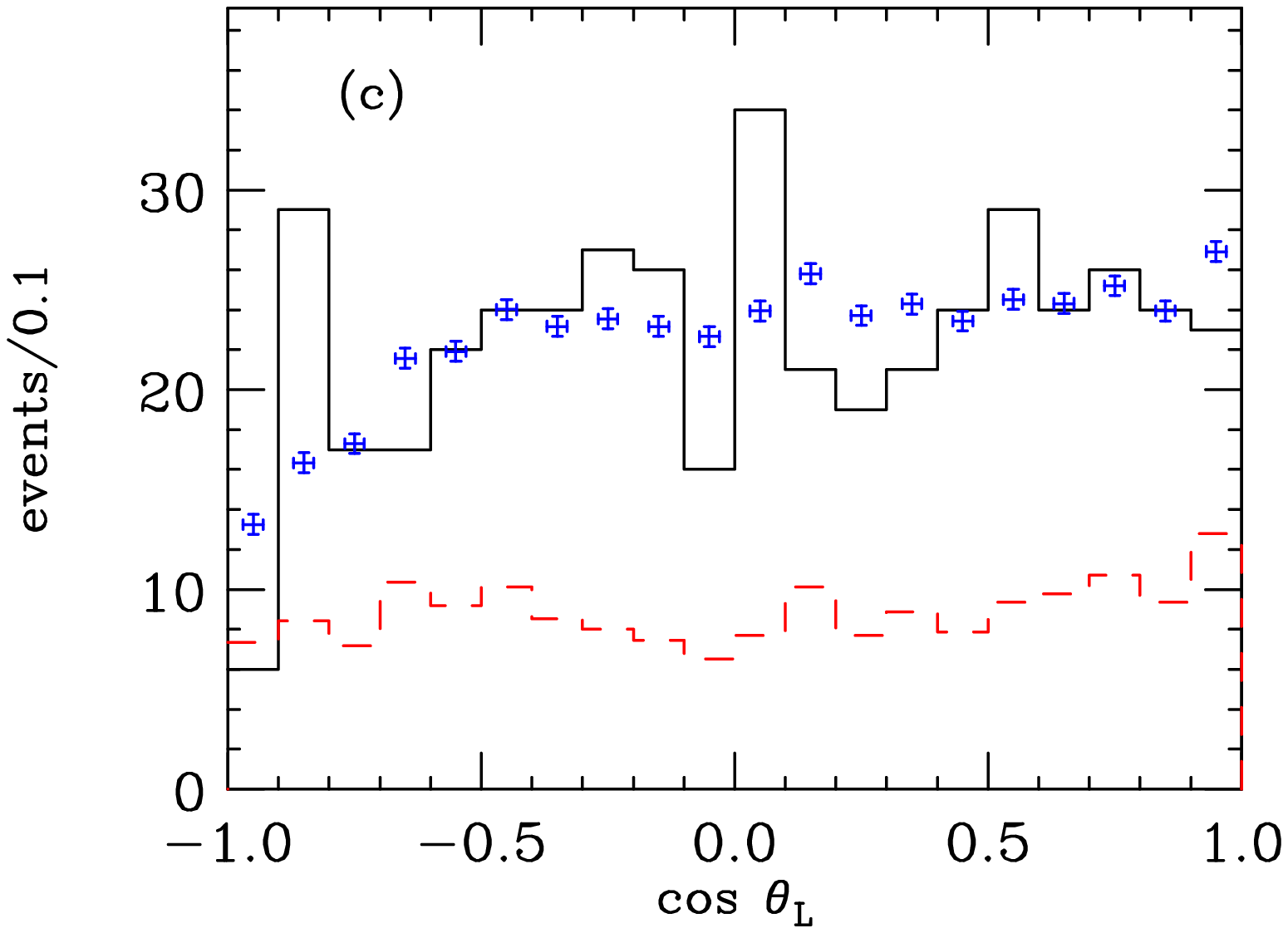}
\end{center}
\caption{Various \costhv{} and \costhl{} projections.  
The data are given by the upper histogram.
The model (crosses with flats) includes the signal computed with the fitted form factor
ratios plus the sideband normalized ccbar background. The lower histogram (dashed) shows
the projections of the ccbar background. 
(a) The \costhv{} distribution for \qsq{}/$\qsq{}_{\rm max} <
0.5$ (b) The \costhv{} distribution for \qsq{}/$\qsq{}_{\rm max} >
0.5$ (c) The \costhl{} distribution for \qsq{}/$\qsq{}_{\rm max} <
0.5$ (d) The \costhl{} distribution for \qsq{}/$\qsq{}_{\rm max} >
0.5$
\label{CVCL}}
\end{figure}

Figure \ref{CVCL} compares the  \costhv{} and \costhl{} distribution between the data and our Monte Carlo model for events at high and low \qsq{}.  As $\qsq{} \rightarrow \qsq{}_{\rm max}$ one expects an isotropic distribution in both \costhv{} and \costhl{} since
all three helicity basis form factors become equal. The data presented in Figure \ref{CVCL} match this expectation relatively well.

\section{Form Factor Ratio Systematic Errors}
Three basic approaches were used to determine the systematic error on
the form factor ratios. In the first approach, we measured the stability of
the branching ratio with respect to variations in analysis cuts
designed to suppress backgrounds. In these studies we varied cuts such
as the detachment criteria, particle identification cuts, vertex isolation cuts,
and the out-of-material cut. Sixteen such cut sets were considered.  In
the second approach, we split our sample according to a variety of
criteria deemed relevant to our acceptance, production, and decay
models and estimated a systematic based on the consistency of the form
factor ratio measurements among the split samples. These included computing separate
form factors for particles and antiparticles, and comparing the values obtained
over the full data set to values obtained from a subset (2/3) of the data
in which the target silicon \cite{target} was operational. In the third approach we checked the stability of the branching fraction as we
varied specific parameters in our Monte Carlo model and fitting
procedure.  These included varying the level of the charm background and in the  
size of the kinematic windows used to associate Monte Carlo events with data points 
in the computation of the fitting intensity.
The only non-neglible systematic error was that due to variation in the results over the sixteen cut selections.
Combining all three non-zero systematic error
estimates in quadrature we find \rvresult{} and \rtworesult{}.

\mysection{Summary}

\begin{figure}[tbph!]
 \begin{center} 
 \includegraphics[width=3.in]{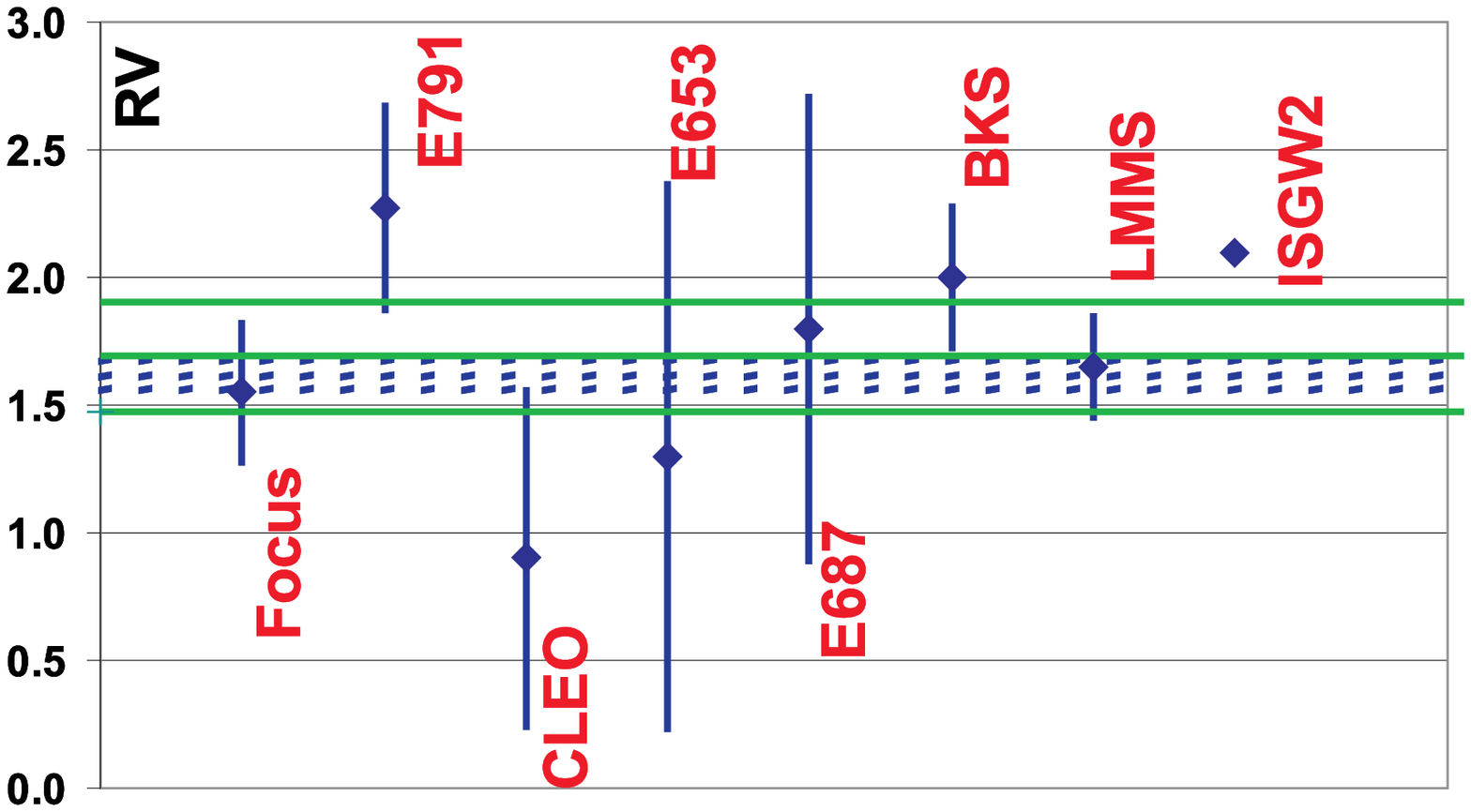}
 \includegraphics[width=3.in]{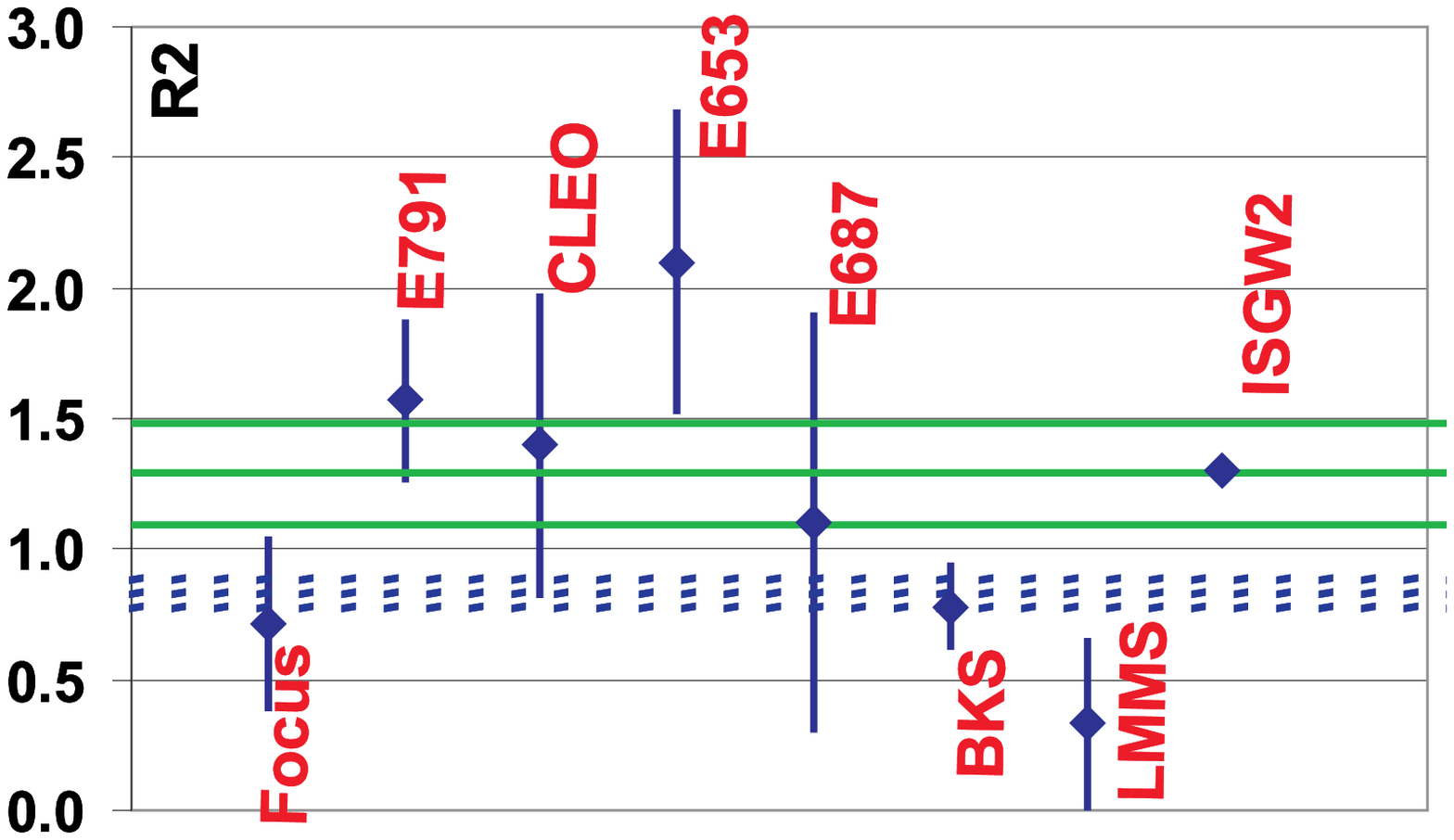}
\caption{ Comparision to previous data and calculations of the \phimndk{} form factors. The three solid green 
lines represent the weighted average of the world's experimental data for \phimndk{} and $\pm 1~\sigma$. The three dashed blue  
lines represent the weighted average of the world's experimental data on \krzmndk{} and $\pm 1~\sigma$.
\label{WA_focus}}
\end{center}
\end{figure}

In Table \ref{ff_table} we compare our results to other experiments. Our
weighted average of all the experimental results is $r_v = 1.678 \pm 0.213$ and $r_2 = 1.292 \pm 0.194$ where systematic errors have been included. We obtain a confidence level of 44.3\% that all 5 experiments have a consistent \rvee{} and a confidence level of 21.5 \% that all \rtwo{} measurements are consistent.  Figure \ref{WA_focus} is a graphical representation of Table \ref{ff_table}.

\begin{center}
\begin{table}[htp]
\caption{Measurements of the \phimndk{} form factor ratios}
\begin{tabular}{l|l|l}
Group & \rvee{} & \rtwo{} \\
\hline \hline
This work & \rvvalue{}  & \rtwovalue{} \\
E791\cite{e791phi}& $2.27 \pm 0.35 \pm 0.22$ &  $ 1.570 \pm 0.250 \pm 0.190$  \\ 
CLEO\cite{cleophi} & $0.9 \pm 0.6 \pm 0.3$ & $1.400 \pm 0.500 \pm 0.300$  \\ 
E653\cite{e653phi} & $2.3 \pm 1.0 \pm 0.4 $  & $2.100 \pm 0.550 \pm 0.200$  \\ 
E687\cite{e687phi} & $1.8 \pm 0.9 \pm 0.2 $ &  $ 1.100 \pm 0.800 \pm 0.100$ \\ 
\end{tabular}
\label{ff_table}
\end{table}
\end{center}

Our measured \rvee{} and \rtwo{} values for \phimndk{} are very consistent with our measured \rvee{} and \rtwo{} values for \krzmndk{} \cite{focus_ff}. The measurements reported here call into question the apparent inconsistency between \rtwo{} values the \krzlndk{} and \philndk{} form factors present in previously published data and are consistent with the theoretical expectation that the form factors for the two processes should be very similar.

\mysection{Acknowlegments}

We wish to acknowledge the assistance of the staffs of Fermi National
Accelerator Laboratory, the INFN of Italy, and the physics departments
of the collaborating institutions. This research was supported in part
by the U.~S.  National Science Foundation, the U.~S. Department of
Energy, the Italian Istituto Nazionale di Fisica Nucleare and
Ministero dell'Universit\`a e della Ricerca Scientifica e Tecnologica,
the Brazilian Conselho Nacional de Desenvolvimento Cient\'{\i}fico e
Tecnol\'ogico, CONACyT-M\'exico, the Korean Ministry of Education, and
the Korean Science and Engineering Foundation.

\end{document}